\documentclass[sigconf]{acmart}

\usepackage{caption}
\usepackage{subcaption}
\usepackage{graphicx}

\usepackage{draftwatermark}
\SetWatermarkText{PRE-PRINT}
\AtBeginDocument{%
  }

\copyrightyear{2025}
\acmYear{2025}
\setcopyright{cc}
\setcctype{by}
\acmConference[CHI '25]{CHI Conference on Human Factors in Computing Systems}{April 26-May 1, 2025}{Yokohama, Japan}
\acmBooktitle{CHI Conference on Human Factors in Computing Systems (CHI '25), April 26-May 1, 2025, Yokohama, Japan}\acmDOI{10.1145/3706598.3713894}
\acmISBN{979-8-4007-1394-1/25/04}

\begin{document}

\title{Exploring the Needs of Practising Musicians in Co-Creative AI Through Co-Design}

\author{Stephen James Krol}
\orcid{0000-0002-9474-3838}
\affiliation{%
  \institution{SensiLab, Monash University}
  \city{Melbourne}
  \country{Australia}
}
\email{stephen.krol@monash.edu}

\author{Maria Teresa Llano Rodriguez}
\affiliation{%
  \institution{University of Sussex}
  \city{Brighton}
  \country{United Kingdom}}
\email{teresa.llano@sussex.ac.uk}

\author{Miguel Loor Paredes}
\affiliation{%
  \institution{Monash University}
  \city{Melbourne}
  \country{Australia}
}
\email{miguel.loor@monash.edu}

\renewcommand{\shortauthors}{Krol et al.}

\begin{abstract}
Recent advances in generative AI music have resulted in new technologies that are being framed as co-creative tools for musicians with early work demonstrating their potential to add to music practice. While the field has seen many valuable contributions, work that involves practising musicians in the design and development of these tools is limited, with the majority of work including them only once a tool has been developed. In this paper, we present a case study that explores the needs of practising musicians through the co-design of a musical variation system, highlighting the importance of involving a diverse range of musicians throughout the design process and uncovering various design insights. This was achieved through two workshops and a two week ecological evaluation, where musicians from different musical backgrounds offered valuable insights not only on a musical system's design but also on how a musical AI could be integrated into their musical practices.
\end{abstract}





\maketitle

\section{Introduction}
Generative music has seen significant advancements in the last 20 years, resulting in modern systems that can generate human-like music in a variety of different genres and mediums \cite{huang2018music,copet2024simple,sturm2015folk,thickstunanticipatory}. These generative systems are framed on a spectrum between end-to-end composition tools \cite{Ahmed_2024,huang2018music} and co-creative tools to aid in human composition \cite{ryan2020,roberts2018hierarchical,tchemeube2023evaluating}.  
To date, the generative capability of modern musical models has shown potential to contribute to complete musical projects \cite{YACHT,music-transformer-album,huang2020ai}, assist in exploration/ideation \cite{MusicVAE,pati2019learning} and to enable better collaboration between less experienced composers \cite{suh2021ai}. Recent work has also demonstrated the value of including practising musicians in the development \cite{deruty2022development} and application \cite{huang2020ai} of these technologies, highlighting an intuitive need for researchers to integrate musicians into the design process. However, much of this work often requires musicians to use an already developed tool and take part in discussions on their creative process with the tool \cite{deruty2022development,huang2020ai,ryan2020}, overlooking the opportunity to engage practising musicians throughout the design process. Music performance and music composition are fields of practice that can take many years to master \cite{EricssonK.Anders1996TAoE,zhukov2009effective}, and while researchers developing co-creative musical AI likely have a musical background, the majority probably do not spend most of their time as professional musicians or actively practising the art of music.  Even if the researcher is an experienced, practising musician, their inherent bias as a technology researcher/creator could result in the development of unwanted tools that have unintended consequences, a common issue in science and technology \cite{koepsell2010genies,Coiera2011}. Therefore, there is motivation to incorporate our target group, in this case practising musicians, throughout the design process, so we can better avoid these issues, as demonstrated in other fields of study \cite{Weitz2024,holloway2019making,Li2023}. 

In this paper, we present a case study that aimed to investigate the needs of practising musicians in co-creative AI through the design and development of an AI-based tool for composition using a co-design methodology. We define a practising musician as a person who \textit{actively} engages in the making and/or performing of music, either professionally or as a hobbyist. Practising musicians can come from different backgrounds and are distinguished from novice musicians who possess some musical experience but rarely engage in the practice of music. Examples of practising musicians are composers, producers, teachers and local artists. This co-design approach led to not only the creation of a co-creative tool with a clear role in our participants' practice, but also uncovered design insights that provided stepping stones to better understand the needs of our practising musicians and how to design co-creative AI systems that are tailored to them. The study involved a total of (n=13) practising musicians with diverse musical backgrounds and consisted of two workshops and a two-week ecological study to design a musical variation tool and explore their individual practices. The contributions of this work are as follows:

\begin{itemize}
    \item \textbf{Co-Design Case Study for Building Co-Creative Musical AI}: We demonstrate how co-design can be used to build co-creative AI for musical practice, and how this methodology can be employed to identify a role for a co-creative AI early in the design process using co-design workshops and an ecological study.
    \item \textbf{Design Insights for Building Musical Co-creative Systems}: Through analysis of our qualitative data, we derived a set of insights that future designers could consider when developing co-creative systems for practising musicians. These insights include the importance of building tools that allow musicians to own their creative process, the effect of framing the technology as a collaborator, and the influence of different musical backgrounds on preferred features.
\end{itemize}

\section{Background}

\subsection{Human-Centred Co-Creative AI}
Co-creative AI systems are demonstrating potential in various fields such as music \cite{ryan2020,MusicVAE,tchemeube2023evaluating,ens2020mmm,huang2020ai}, writing \cite{gero2023social,wang2024}, drawing \cite{lin2020your,oh2018lead,zhang2021} and design \cite{rajcic2024towards,Zhou2024}. Several frameworks for designing co-creative systems have been proposed \cite{Moruzzi2024,Zhou2024,Rezwana2023}, reflecting the increasing interest in refining approaches to the development of co-creative AI. These frameworks present a multifaceted approach with objectives ranging from understanding the roles and varying levels of contribution that co-creative systems can offer in human-AI interactions \cite{lubart2005can,guzdial2019friend}, the implications of these roles \cite{buschek2021nine}, different collaboration modalities \cite{kantosalo2020modalities,Rezwana2023,Moruzzi2024}, and specific aspects of the interaction \cite{llano2022explainable}. Additionally, there have been several studies aimed at understanding what users need from these systems. Oh et al. (\citeyear{oh2018lead})\cite{oh2018lead} explored the user experience of participants using the co-creative drawing system DuetDraw, highlighting insights such as the importance of offering clear instructions on system usage and enabling users to take the lead in the creative process. Similarly, Santo et al. (\citeyear{santo2023focusing})\cite{santo2023focusing} used a probing tool, POCket Artist (POCA), to explore general artist needs for creative software, emphasising the importance of artist-centered development in this domain. Through the development of a co-creative tool named Cococo, Louie et al. (\citeyear{ryan2020}) investigated the needs of novice musicians with regards to musical AI-steering tools. The study included an initial needs assessment with novice musicians that identified several issues with both a deep generative model and its interface. This was then used to inform the design of their Cococo system which was subsequently evaluated by 21 musical novices. This evaluation demonstrated an improvement in the system and resulted in various design insights such as the value of AI transparency and the importance of building semantically-meaningful tools. In our work, we aim to understand the needs of practising musicians in co-creative AI, who, due to their more complex skill set and advanced understanding of music, are likely to have distinct needs across different systems compared to visual artists \cite{oh2018lead} and novice musicians \cite{ryan2020}. 

\subsection{AI in Music: Evaluation and Design Methodologies}
\subsubsection{Music AI Evaluation Methodologies}
Various deep learning models have been developed to perform a range of musical tasks, including generating entire pieces or phrases \cite{huang2018music,performancernn,mittal2021symbolic,copet2024simple}, in-painting within an existing composition \cite{MusicVAE,ens2020mmm,thickstunanticipatory,huang2019bach}, and musical style transfer \cite{wu2023musemorphose}. Recent advancements in multi-modal models \cite{radford2021learning,rombach2022high} have also enabled text-to-music generation \cite{copet2024simple,agostinelli2023musiclm}, with new private companies now offering complete song generation services\footnote{https://suno.com/}. With this, there has been development of models with inbuilt control mechanisms that can be used as co-creative tools \cite{pati2019learning,pati2021attribute,bryan2024exploring,TransformerVAE,wu2023musemorphose,roberts2019magenta}. However, because these systems are approached as machine learning problems, their evaluation is often restricted to beating quantitative benchmarks and conducting basic human listening tests, providing little understanding on their usefulness as tools to musicians. Louie et al. (\citeyear{louie2022expressive}) \cite{louie2022expressive} highlight the disparity in how co-creative musical systems are evaluated within the fields of Machine Learning (ML) and Human-Computer Interaction (HCI), and argue that both methodologies are limited, with ML metrics serving merely as proxies, while HCI methods are overly subjective and fail to adequately address the final artefact. The authors constructed their own evaluation method that involved both a subjective assessment from the user as well as outsider listening tests to evaluate the quality of the generated artefacts. Tchemeube et al. (\citeyear{tchemeube2023evaluating}) \cite{tchemeube2023evaluating} also introduced a novel methodology for evaluating Human-AI interaction through usability, user experience and acceptance measures and utilises both quantitative and qualitative techniques to provide an all-round assessment of the tool. While both methodologies are suitable, they focus on the evaluation of an already developed system. In contrast, our work focuses not on evaluating a co-creative system, but on using a co-design approach to develop a tool and gain insights into the needs of practising musicians. 

\subsubsection{Music AI Design Methodologies with Practising Musicians}
Although significant progress has been made in developing co-creative musical AI \cite{MusicVAE,ens2020mmm,banar2023tool,pati2021attribute}, less work has focused on involving practising musicians in the design and development process \cite{deruty2022development}. Instead, many works utilise common musical practices \cite{banar2023tool,thickstunanticipatory} to infer the needs of their target group and build models that support these processes. For example, Thickstun et al. (\citeyear{thickstunanticipatory})\cite{thickstunanticipatory} observed that many generative transformer models produce music in a linear, start-to-finish manner. However, human composition is often non-linear and follows a more iterative approach. To address this, they designed their anticipatory transformer to use both previous and future notes when infilling a composition, better complementing the human composition process. However, while this methodology has produced valuable technologies, it risks overlooking less obvious needs of musicians, as demonstrated by Huang et al. (\citeyear{huang2020ai})\cite{huang2020ai} in their study on the AI Song Contest, which identified several challenges impacting the overall usability of AI tools.

To address this, researchers have employed various methods to include musicians in their work; for example, Google's Magenta have worked with musicians to create complete projects \cite{YACHT,music-transformer-album} and documented their experiences from these collaborations. Deruty et al. (\citeyear{deruty2022development})\cite{deruty2022development} conducted a participatory design study with six practising musicians, who provided regular feedback to evaluate and improve a set of AI tools developed by the research group, generating various design insights through these interactions. Similarly, Ford et al (\citeyear{ford2024}) \cite{ford2024} requested 6 composers to reflect and record their impressions of using different AI tools, aiming to contribute a novel method for creative support tool research. 
Other studies have taken advantage of the researcher's own experience as a musician to gather insights; for example, Benford et al (2024) developed an autonomous system with the ability to perform a duet with musicians through a practice-based study, collecting insights related to generating trust and negotiating autonomy in real time interactions. In an autoethnographic study, Sturm (\citeyear{sturm2022generative}) \cite{sturm2022generative} documented his experience, and explored the benefits of generative AI systems for co-creation in songwriting. His autoethnography highlights various opportunities despite the unpredictability and lack of control in the results from these ‘creative partnerships’. 

While these works highlight the value of collaborating with practising musicians, more can be done to improve how musicians are involved and ensure that perspectives from groups less familiar with AI are also considered. Rather than infer the needs of users \cite{banar2023tool,thickstunanticipatory,ens2020mmm,MusicVAE} our work uses a co-design approach from the outset of development to identify the needs of practising musicians throughout the design process. Our work is distinguished from previous work \cite{huang2020ai,deruty2022development,YACHT,music-transformer-album} in that we use a more rigorous participatory design methodology to develop our tool and gain insights on our user group.

\subsubsection{Co-design as an approach for Music AI Development}
Participatory Design (PD), an approach that aims at actively involving relevant stakeholders in the process of design, has been reported in the field of human-computer interaction since the 1980s \cite{sanders2008, sarmiento2022}. Associated to PD, there is an umbrella of methods in which stakeholders have some kind of participation during the design, some of which include: Co-creation, User Centred Design, Collaborative Design, and Co-Design. Although often in the human-computer interaction literature these methods are used interchangeably \cite{hi2015comparative, grindell2022use, McGill2022, sarmiento2022}, our work is guided by the principles of Co-Design. 

A key distinction of co-design, compared to methodologies like User-Centered Design (UCD), is that it views the user as a partner rather than merely a subject or source of information \cite{sanders2008, robertson2012participatory, hi2015comparative}. In this respect, Co-Design calls for user participation rather than user evaluation, reflecting their experience and knowledge throughout the design process rather than just interpreting their needs \cite{rizzo2011}. We follow this philosophy through the implementation of two co-design workshops and an ecological study. Instead of focusing on typical test and feedback sessions, we aimed at creating opportunities for participants to situate themselves within their practice through composition tasks and actively take part in designing the functionality of a co-creative AI. This not only resulted in the development of an AI system, but also led to uncovering aspects of their creative process that impact their interactions with co-creative AI.

Although co-design has been used in multiple domains\footnote{A quick Google Scholar search for the term 'Co-design', reveals a prevalent use in the health domain, while a search for the terms ```Co-design'' + ``Music''', reveals a key focus on the design of instruments targeted to people with impairments.}, to the best of our knowledge, the use of co-design with practising musicians is scarce and limited. Our work contributes to the field with a case study on the use of co-design with practising musicians, revealing key insights for the development of co-creative AI for music and resulting in the development of a variation tool.

\section{Case Study: Co-Designing a Deep Learning Based Music Variation Tool}

\begin{figure}[h]
    \centering
    \includegraphics[width=\linewidth]{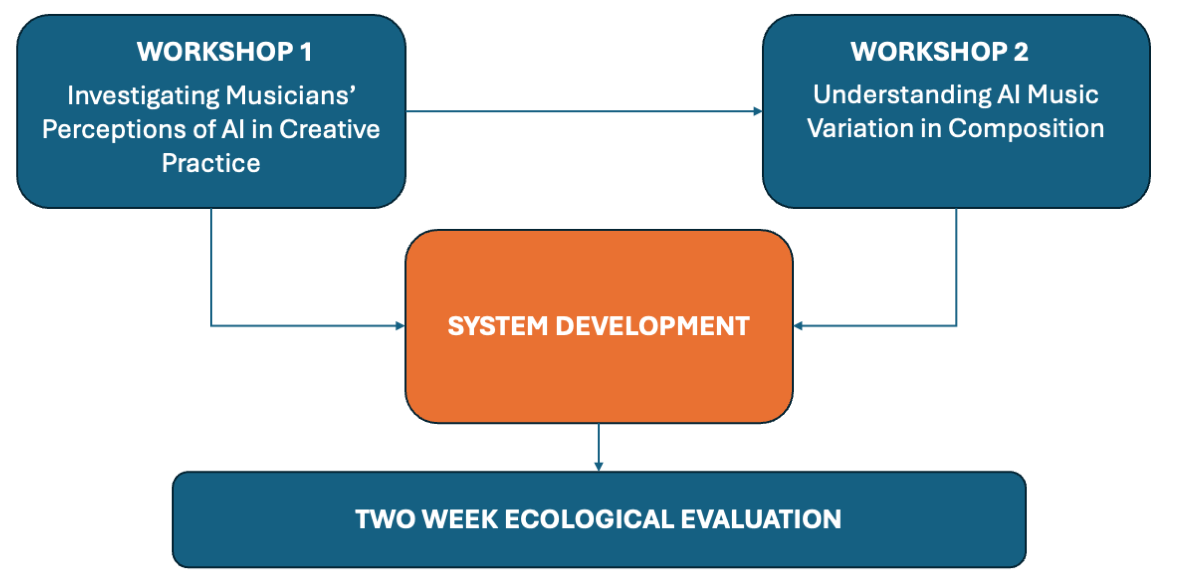}
    \caption{Study workflow: Workshop 1 was used as the initial investigation into how a co-creative AI can be situated in a musical practice and informed the design of workshop 2. Following insights from the previous investigation, Workshop 2 was focused on understanding the use of a variation system,  and with workshop 1, was used to continue guiding the system development. This system was then used in an ecological evaluation to better understand participants specific needs from the system.
    }
    \Description{Depicts the work flow of the study. A the top there are two boxes next to each other. Box 1 is titled Workshop 1 Investigating Musicians' Perceptions of Ai in creative practice. Box 2 is titled Workshop 2 Understanding AI music variation in composition. There is an arrow leading from Box 1 into box 2 signifying that workshop 1 influenced workshop 2. Both box 1 and 2 have an arrow leading to a lower box titled system development indicating these workshops influenced how the system was built. This box then points to a final box titled two week ecological evaluation.}
    \label{fig:study-workflow}
\end{figure}

To explore how a co-creative AI could be integrated into the practice of musicians, we collaborated with (n=13) practising musicians to develop a co-creative tool. Similar studies have successfully used this method \cite{ryan2020,deruty2022development} in creative research, with co-design as a whole being a well-established methodology for designing technology \cite{Weitz2024,Nicholson,Harrington2018}, providing a good justification for our approach. To qualify for the study as a practising musician, participants needed to actively engage in the making and/or performing of music. This could be as a professional or as a hobbyist and was not limited to a specific musical background.

Two workshops were held to create a starting point for development. The first workshop investigated the general perceptions of co-creative AI from a group of practising musicians who identified an AI tool that could fit in their practices. The second workshop further explored this system and provided functionalities for the tool that would be needed for it to be useful in a musical practice. The insights from both workshops guided the development of a music variation tool that musicians could use for ideation. To gather deeper insights and further design suggestions, we conducted a two-week ecological evaluation in which participants used the system in their own environments. They then took part in a focus group to discuss how this tool, and co-creative AI more broadly, could integrate into their practices and to suggest additional functionality they desired in the tool. The following section outlines the case study in detail.

\subsection{Workshop 1: Investigating Musicians’ Perceptions of AI in Creative Practice}
As an initial step in exploring how a co-creative AI could be designed with practising musicians in mind, we conducted a workshop to gather insights into their perceptions of AI and to identify potential roles a co-creative system could play in their creative practice. A total of (n=5) musicians took part in the workshop with all participants possessing formal training in music. Participants were recruited through both the university's music faculty and the lab's known network of musicians. Demographic information in regards to each participant can be seen in Table \ref{tab:w1participants}.

\begin{table}[h]
    \centering
    \resizebox{\columnwidth}{!}{
        \begin{tabular}{|c|c|c|c|}
        \hline
         Participant & Gender & Age & Background  \\
         \hline
         P1  & M & 25-34 & Classical pianist and member of local band. \\
         P2 & F & 18-24 & Classical background in composition. \\
         P3 & M & 18-24 & Contemporary background in composition. \\
         P4 & M & 34-45 & Jazz Performance. \\
         P5 & M & 34-45 & Classical french horn player and musical educator. \\
         \hline
        \end{tabular}
    }
    \caption{Demographic information of participants who took part in Workshop 1.}
    \Description{Contains demographic information for participants in workshop 1. There are are a total of 5 participants with information on their gender, age and musical background.}
    \label{tab:w1participants}
\end{table}

Before the workshop, participants were asked to compose complete pieces and to take note of their composition process. The workshop was conducted over a 3 hour period and split into three sections. (1) A discussion on their composition process and understanding how they make decisions, (2) an active composition session where participants took turns composing music with either another participant or an AI agent and (3) a final discussion on the role AI could play in their creative practice and their overall views on the potential of AI in music. Participants were provided with a \$80 USD equivalent gift card for their time.
The data collected from the study was qualitative, consisting of transcripts from the discussions and video recordings of the composition sessions. 

All transcript data collected in this study was analysed using an inductive thematic analysis \cite{braun2006using}. This involved two researchers independently coding the same data and identifying themes. Researchers then engaged in a discussion, comparing their codings to determine a final set of codes and themes consistent with each others findings. By independently coding the same data, we hoped to mitigate individual researcher bias and better identify the themes present in the data.

\subsubsection{Understanding How Participants Approach Composition}
This part of the workshop was essential in understanding our participant's identity as musicians and composers and provided a strong foundation for the discussions held later in the workshop. The composition task set prior to the workshop also provided set pieces that each participant could refer to when explaining their process, giving a unique insight into how they made music. 
From this we built an understanding of shared techniques used by the participants such as \textit{Composing from a Theme} which involved using a concept - "for me, I always like to stick to the concept because I feel like that's what grounds me" (P2) or emotion - "I had a feeling that I was trying to convey and that was conveying that feeling for me" (P1) or imagery - "the process is actually more have an emotion or an image or an experience, very visual sort of thing" (P5) to drive composition. This discussion also highlighted the importance of \textit{Musical Intuition} on composition and how it frequently led to decisions that could only be articulated in terms of the emotional response they evoked in the composer, with one participant stating - "how could you explain something that comes from the heart rather than from your logical thinking?" (P3).
Ultimately, many discussion points led to the notion of \textit{Human Embodiment} and how much of the music was driven by their experience as humans - "It was a rainy shitty day. And I was just like, oh, couldn't be bothered as I go. So really slow chords. This like Oh, that's just how it feels" (P5) and the craft they have developed over many years - "there's actually a lifetime of perfecting a craft behind this and you can only sort of scratch the surface of what it's really about" (P4). These discussion points became especially relevant later in the workshop when AI was introduced into the mix.

\subsubsection{Exploring Human-Human and Human-AI Composition Through Play}
After the initial discussion, participants engaged in a composition activity where they had to compose a series of short compositions from randomly selected prompt cards. These prompt cards contained simple scenarios that the participants could use as inspiration for their compositions. Examples of provided prompts included "You come home and are greeted by an excited puppy" and "A tarantula unexpectedly falls into your lap". Prompting has been used in other studies to evaluate co-creative systems \cite{ryan2020,louie2022expressive} and was positively received by our participants. 
In addition to the prompt cards, participants had to either compose with another participant or with an AI generative system. The AI system was powered by Music Transformer \cite{huang2018music} and utilised a ported script \cite{piano_transformer} to build an application for users to interface with the model. The interface was simple and allowed users to input MIDI \footnote{MIDI is a symbolic format for representing digital music where notes and other musical features are formatted as events} live through a MIDI keyboard and generate re-harmonisation's of their music. Music Transformer was selected due to its impressive generative capabilities, often attributed to the large 10,000 hour midi dataset it was trained on \cite{music-transformer-data}. The purpose of this activity was not only to provide participants with the chance to interact with an AI system but also to offer them a basis for comparing the experience of creating with AI versus creating with another human. This approach enabled us to initiate discussions on what it would mean for the participants if AI were to function not merely as a tool but also as a collaborator.

\begin{figure*}
    \centering
    \begin{subfigure}[b]{0.3\textwidth}
         \centering
         \includegraphics[width=\textwidth]{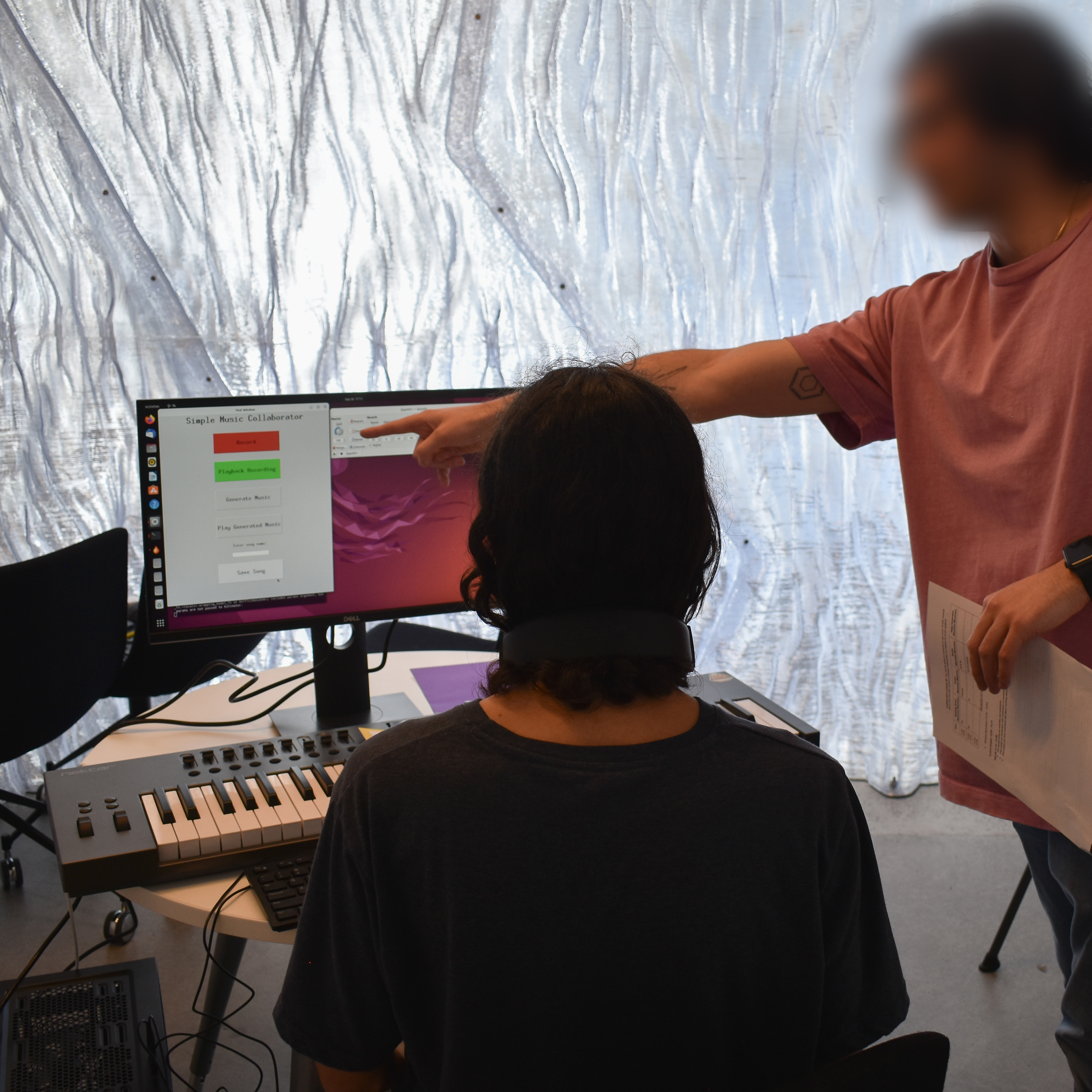}
         \caption{}
         \label{fig:Human-AI1}
    \end{subfigure}
        \begin{subfigure}[b]{0.3\textwidth}
         \centering
         \includegraphics[width=\textwidth]{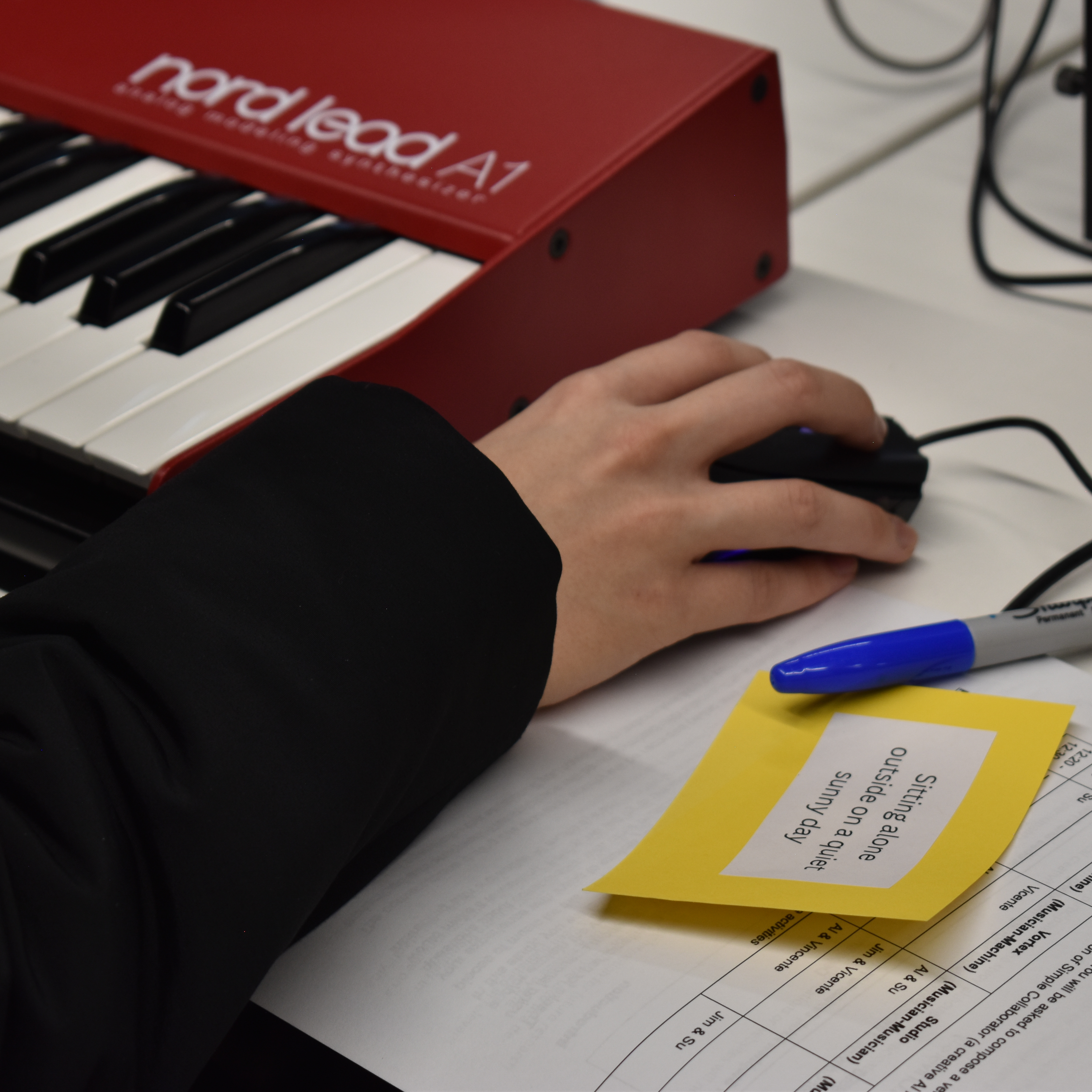}
         \caption{}
         \label{fig:Human-AI2}
    \end{subfigure}
        \begin{subfigure}[b]{0.3\textwidth}
         \centering
         \includegraphics[width=\textwidth]{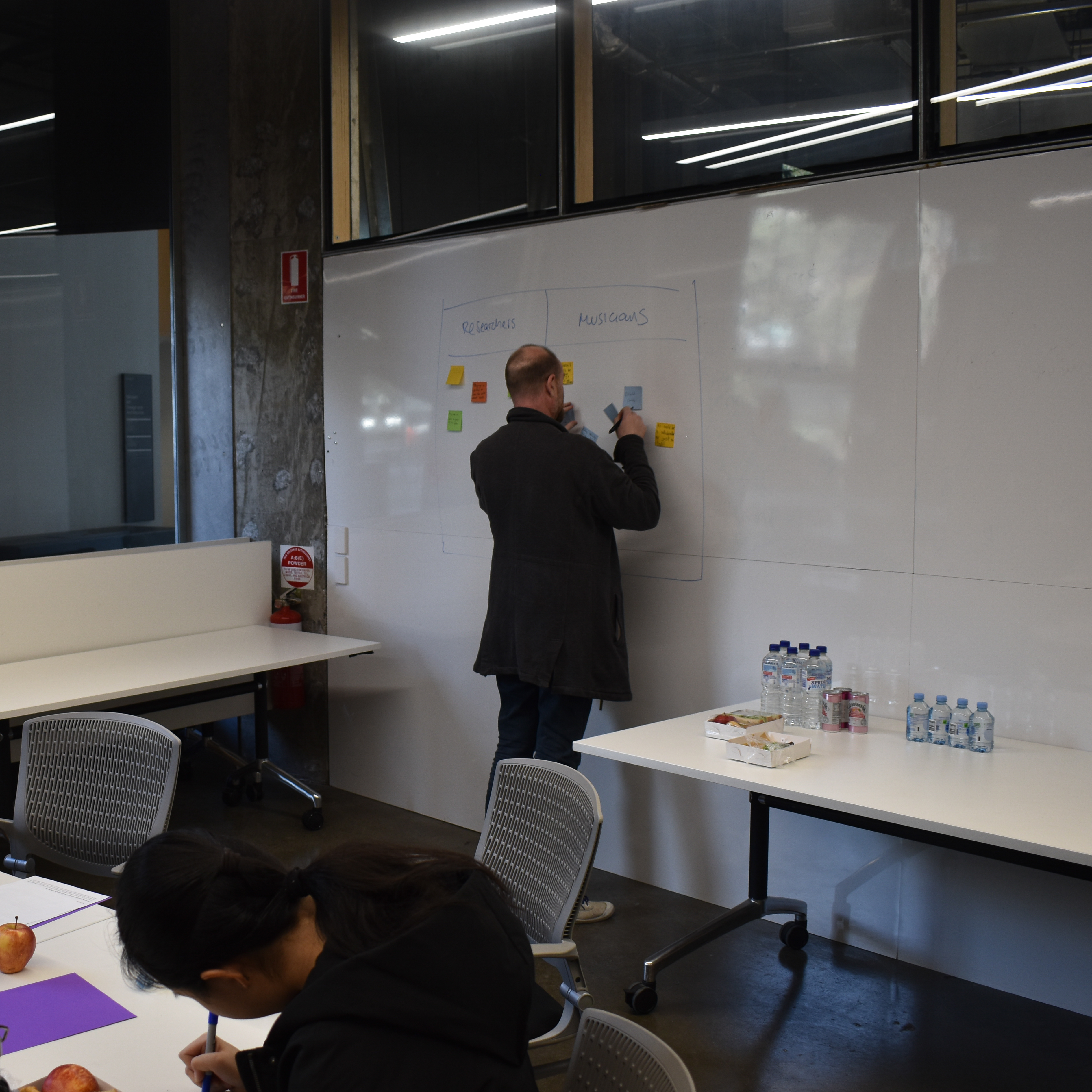}
         \caption{}
         \label{fig:v}
    \end{subfigure}
    \caption{Figure a: Musician being briefed on the musical system and preparing to compose with it. Figure b: Example of prompt card used by musicians when composing. Figure c: Picture from the workshop discussions with a participant placing ideas on the board.}
    \Description{Contains three images from workshop 1. Image A depicts a participant with an AI system and a MIDI keyboard. Image b depicts a prompt card next to a musician. Image C depicts a participant adding ideas to a whiteboard.}
\end{figure*}

\subsubsection{Defining AI's Position in Musical Composition} 
Following the composition task participants engaged in a discussion about AI's future role in music creation and the potential for AI to be seen as not just a tool but as a collaborator. Data from this discussion was used to not only influence the design of our current system but also the framing of the system. This was due to strong push back on the concept of \textit{AI as a Collaborator} which in many cases was due to \textit{AI lacking human experience}. To the participants, music was a means of communication, with P4 stating "it's about empathy, really, and being able to share other people's emotions and experiences and AI has no empathy or understanding", the importance of empathy was also echoed by other participants "I think you actually need empathy, to collaborate, you need that human emotion that the AI can't have." (P3) and "one last thing about that empathy thing. I think we need to recognise that AI can never truly understand what we're hoping for" (P5). Participants also expressed hesitation about collaborating with a machine, as they wanted to preserve \textit{Ownership of the creative process}. Strong sentiments regarding this theme were shared by the participants "There are times when I think writing the melody or the chords, that's ours, like, sorry, but the creative process, it's, that's our possession, that's what we do." (P5), "And I probably wouldn't use a system because, a lot of like composing music for me is actually that process of thinking, what do I want" (P1). 

Interestingly, participants were ok with sharing the creative process with another human "No, no, I think the machine don't have the right to have an opinion. And the human totally has the right to say so 'oh, you disagree with me?' Awesome. let's try to do something else, but with a machine it's like, No" (P3). In fact, sentiments were mostly positive when talking about \textit{Human-to-human Collaboration} and it was clear that this was considered special to the participants - "Different people's ideas coming [it], just strengthens us. It's just a much richer process, the collaboration, because you do have that diversity of thought." (P5), "The inspiration and energy is like 1,000\% more with the other people" (P4) - with the only drawback being that "you have to compromise" (P5). When discussing AI collaboration, participants would often refer to their \textit{Musician Values} "a large part of why we do what we do is because we enjoy that feeling of writing music, of translating ideas into music" (P5), "when you're not creating music as a commodity, that's 100\% of the value was the fun times when you're making it with someone else" (P4), with one participant stating "At what point does a person stop being a composer?" (P5), highlighting the personal significance of the participant's identity as composers. 

Despite resisting the idea of an AI musical collaborator, participants did offer insights into what would make a machine collaborator more attractive with the main points related to \textit{Familiarity}. P5 stated "it's about time, it's about safety. It's about getting the sense that we have input that we can communicate that I will be heard, that they are listening" with P2 adding that "these things might feel more natural" for future generations who will grow up with AI technology. However, participants were generally more open to using AI as a tool "I think with AI, you shouldn't let go of that ego, I think AI shouldn't be a collaborator...it's more of a tool, or completely a tool" (P3) and importantly that this tool should be \textit{Controllable} with participants wanting a tool to have \textit{Varying Agency} - "if you can slide from AI takes over and interprets purely, versus AI takes little bits of it" (P5). On the discussion of AI as a tool, participants often spoke about the potential of a tool that could generate \textit{Musical Variations}. This tool could be used to give "another direction, maybe give sets of options" (P3) and was likened to PowerPoint's slide design tool which adds "different colours and things...but you can still see it's the same structure" (P2). Participants were generally receptive to this idea with P5 stating "I like that. Personally, for me, that would be brilliant" noting that "once you've got the basic ideas [I] could flesh it out, or add variation...it would be good to have some ideas".   
\subsubsection{Workshop 1 Outcomes}
Although some insights shared by participants align with findings from other general co-creative studies—such as the importance of controllability in co-creative tools and varying agency \cite{oh2018lead,ryan2020,Amershi2019}—the workshop still uncovered new perspectives on where a co-creative AI would be most beneficial in their \textit{musical} practice. For example, it was clear these musicians were not looking for a machine collaborator nor an ideation tool that would perform most of the work. They desired to own their creative process and were more interested in a co-creative system that could help expand their current musical ideas. This is not to suggest that AI collaborators are generally undesirable to practising musicians; however, the consensus among participants in this study indicated that how we frame this technology, for instance, either as a collaborator or a tool, influences its adoption within practising musicians. It was also clear that participants valued human-human collaboration and were not interested in replacing this with an AI. Following this workshop, it was decided to focus development on creating a music variation tool, based on the participants' suggestions that such a tool would be valuable.

\subsection{Workshop 2: Understanding AI Music Variation in Composition}

After the initial workshop, a second workshop was conducted to explore: (1) whether a musical variation tool would be useful in musical practice, and (2) what features musicians would want in such a tool. Although this workshop centred on a specific type of system, discussions still led to insights that apply beyond the tool. A total of (n=6) musicians took part in this study, all different from the previous workshop to gain new perspectives on this type of tool. They were also recruited through the university's music faculty and known networks to the researchers. Demographic information relating to participants can be seen in Table \ref{tab:w2participants}.

\begin{table}[h]
    \centering
    \resizebox{\columnwidth}{!}{
        \begin{tabular}{|c|c|c|c|}
        \hline
         Participant & Gender & Age & Background  \\
         \hline
         P6  & M & 25-34 & Professional Composer \\
         P7 & - & 18-24 & Songwriter and Local Performer \\
         P8 & F & 22-34 & Background in Classical Piano\\
         P9 & M & 25-34 & Electronic Instrument Designer \\
         P10 & F & 25-34 & Songwriter and Local/International Performer. \\
         P11 & F & 25-35 & Songwriter, Composer and Professional Producer.\\
         \hline
        \end{tabular}
    }
    \caption{Demographic information of participants who took part in Workshop 2.}
    \Description{Contains demographic information for participants in workshop 1. There are are a total of 6 participants with information on their gender, age and musical background.}
    \label{tab:w2participants}
\end{table}

Workshop 2 differed from Workshop 1 in that it did not include a human-to-human activity, allowing participants more time to engage with an actual AI tool. We believed this approach would better support the design discussions for the variation tool compared to the human-to-human task which was initially designed to understand what an AI would need to act as a collaborator. The pre-workshop activity was also omitted based on feedback from previous participants, who found it overly time-consuming and unnecessary for explaining their composition process. The workshop was held over a three-hour session and was split into two parts: (1) Composing with a prototype music variation system and (2) A discussion on their experience with the system, focusing on whether a tool like this could fit into their practice and what features they would want in it. Participants were compensated with a \$20 USD equivalent gift card for their time. The data collected from the study was qualitative in the form of transcripts from the discussion and was analysed using the same method described in section 3.1.

\begin{figure*}
    \centering
    \begin{subfigure}[b]{0.3\textwidth}
         \centering
         \includegraphics[width=\textwidth]{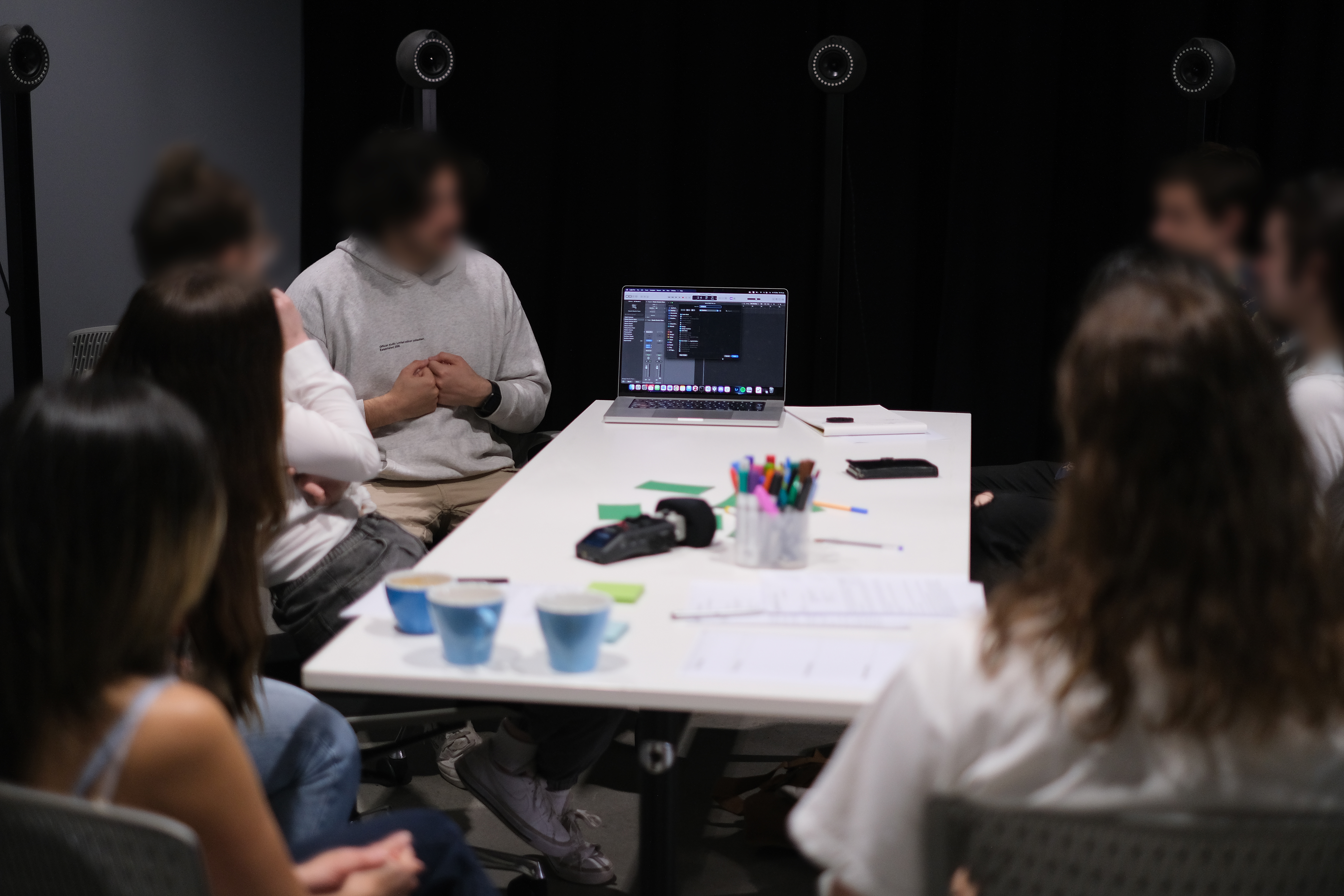}
         \caption{}
         \label{fig:W2-1}
    \end{subfigure}
        \begin{subfigure}[b]{0.3\textwidth}
         \centering
         \includegraphics[width=\textwidth]{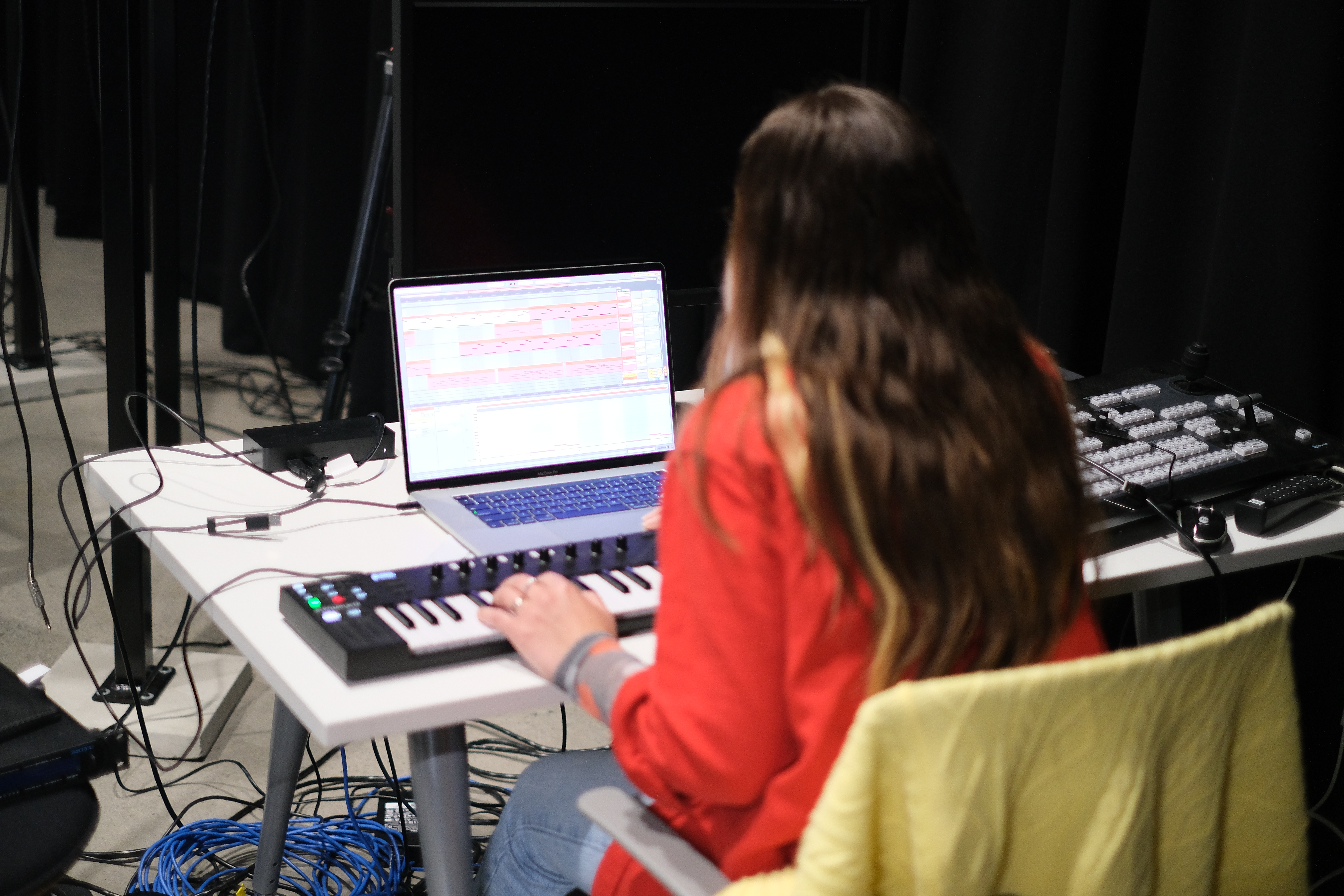}
         \caption{}
         \label{fig:W2-2}
    \end{subfigure}
        \begin{subfigure}[b]{0.3\textwidth}
         \centering
         \includegraphics[width=\textwidth]{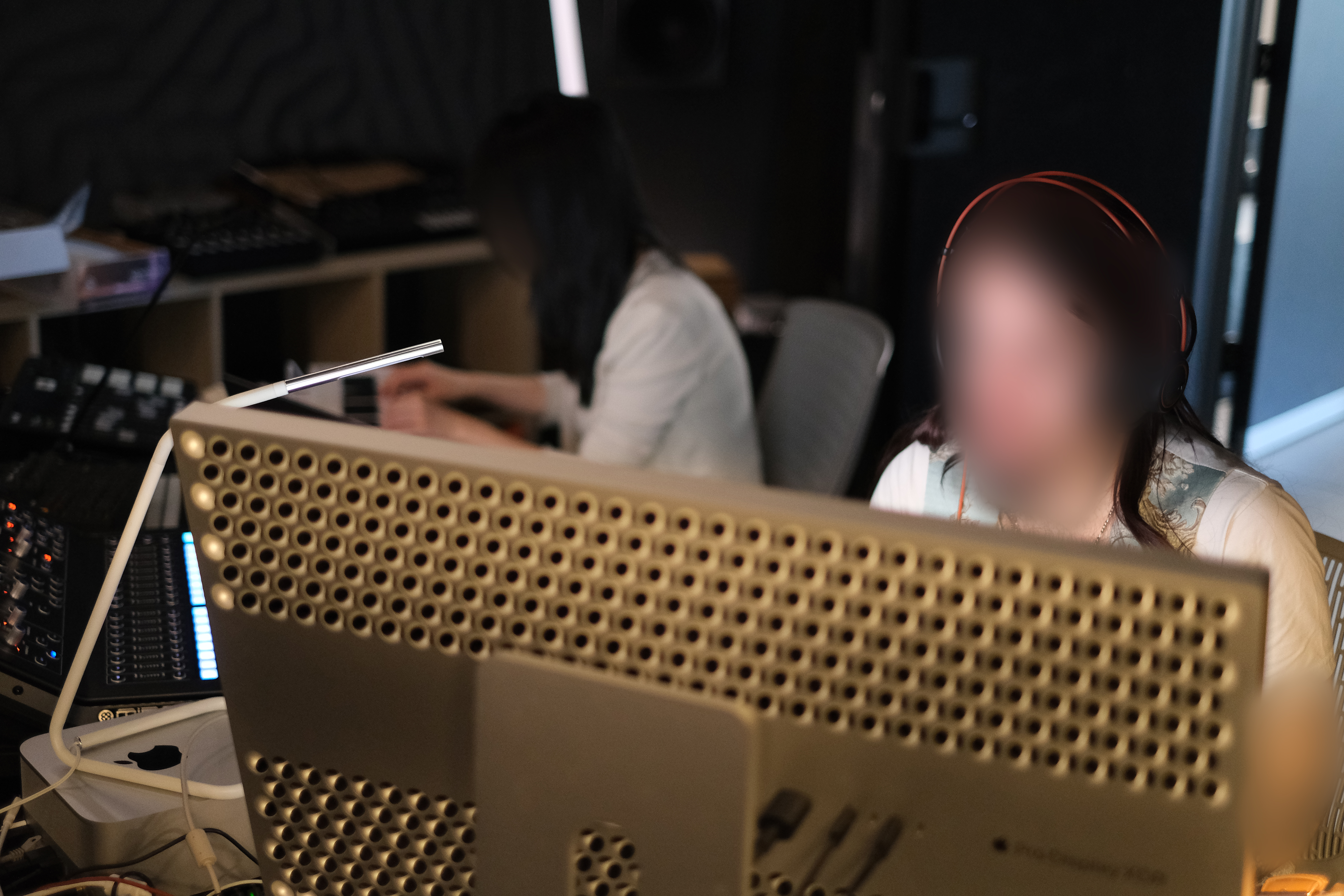}
         \caption{}
         \label{fig:w2-3}
    \end{subfigure}
    \caption{Figure a: Workshop discussion on music variation tools. Figure b: Participant working on a composition with the system. Figure c: Participants composing music in a studio.}
    \Description{Contains three images from workshop 1. Image A depicts all participants in a discussion. Image b depicts a musicians composing with an AI and a view of their DAW. Image C depicts two musicians composing with an AI in a studio.}
\end{figure*}

\subsubsection{Probing Musical Variation Tool}
In this work, we define a music variation tool as a system that modifies musical phrases or pieces based on specific musical attributes. For instance, the tool could re-harmonise a melody \cite{huang2018music}, apply a different rhythmic groove to a pop song \cite{MusicVAE}, or even perform a style transfer by changing the genre of the final piece \cite{wu2023musemorphose}. This concept is well-established in creative research and is often described as 'remixing' \cite{Zhou2024,remixing}. However, in this work, we use the term "variation" due to its historical significance in music \cite{Schwarm_2013} and because it better aligns with this system's functionality. Deep learning was selected as the main engine for this tool due to its current state-of-the-art generative capabilities  \cite{thickstunanticipatory,huang2018music} and its ability to create latent representations, which are well-suited for generating variations \cite{MusicVAE,pati2019learning}. While there exist models for exploring variations \cite{MusicVAE} and discussions on using these tools in composition \cite{banar2023tool}, these models are limited in their generative capabilities, typically being applied only to simple melodies or short, fixed-length bars. Furthermore, to the best of our knowledge, no work has been done to develop these tools specifically for and in collaboration with practising musicians. 
Due to its impressive generative capabilities, the Music Transformer was, once again, used as a probing tool to give participants a reference point when considering a music variation tool. While not specifically designed for music variation, the model does offer a re-harmonisation version which allows users to explore different accompaniments to a specified melody. Participants were provided access to the re-harmonisation model via a simple web-based interface that allowed them to upload MIDI files to a server, which then generated multiple harmonic variations for them. When interacting with the model, participants had the option to either compose songs based on the prompt cards from the previous workshop or to work on their own music.

\subsubsection{Designing a Music Variation Tool}
Following the composition activity, the participants took part in a discussion regarding music variation tools. Through this we gathered more evidence from new participants that a tool like this would be \textit{Useful} in their practice "I would definitely use it in my creative practice and even in my professional jobs" (P11) - "I think it could be really cool, for a lot of the stuff I was making" (P7). Additionally, we found that a variation tool would be especially valuable during the \textit{Ideation} phase. Participants noted that it could "influence new parts of the story" (P7), with one participant sharing that the idea they used "was an idea that I hadn't thought of at all" and that it "became a big structural part of the piece" (P9). 

The effect of ideation varied for different participants. For example, P11 used suggestions from the system to create a "more dissonant and dark chord progression as a B section" for their piece, while P10 focused on a single note introduced by the system, incorporating it into their composition "to capture the idea". This initial discussion provided further evidence indicating a desire to use a musical variation tool in practice, with the primary use case being \textit{ideation}. 

The use of Music Transformer also gave participants a reference point from which they could propose design features for a future system. For example, participants referred to there being \textit{Not Enough Variation} or \textit{Too Much Variation} and noted a desire to \textit{Control the Type of Variation} - "something like a slider to control the level of variation would be helpful to see, if I want less variation or more variation" (P11). There were discussions on a \textit{Personal Model} which could mimic the composers style - "ideally I would like to be able to train it with my own ideas" (P6) and the value of maintaining performative information in the generated MIDI data - "MIDI piano can sound really not human at times, it doesn’t sound like natural organic playing, but what the systems gave back was in this case was just the perfect amount of quantised but like slightly out so it sounded more human" (P11). Participants also suggested other use cases for the system such as packaging it with the prompt cards to create a song writing exercise tool - "something like this could be really helpful for that, like for song writers and composers to just sit and write 30mins of music once a day and build those skills" (P7). Finally, participants shared insights into where they would want to use the system, with the most notable being integrated into their Digital Audio Workstation (DAW) "yeah if it could be implemented within the DAW it would probably be easier for me" (P10).

\subsubsection{Workshop 2 Outcomes}
The second workshop gave us deeper insight into the role a co-creative AI could play in practising musicians' practices and provided evidence that they would be open to incorporating an AI-based system into their workflow. Additionally, the workshop further confirmed that the concept of a musical variation tool was desirable to musicians and highlighted that the tool would be most beneficial during the ideation phase. Finally, it helped identify specific design suggestions, such as control over the level and type of variation - providing clear steps for future development. 

\subsection{MusicBert Variation}

\begin{figure}
    \centering
    \includegraphics[width=0.9\linewidth]{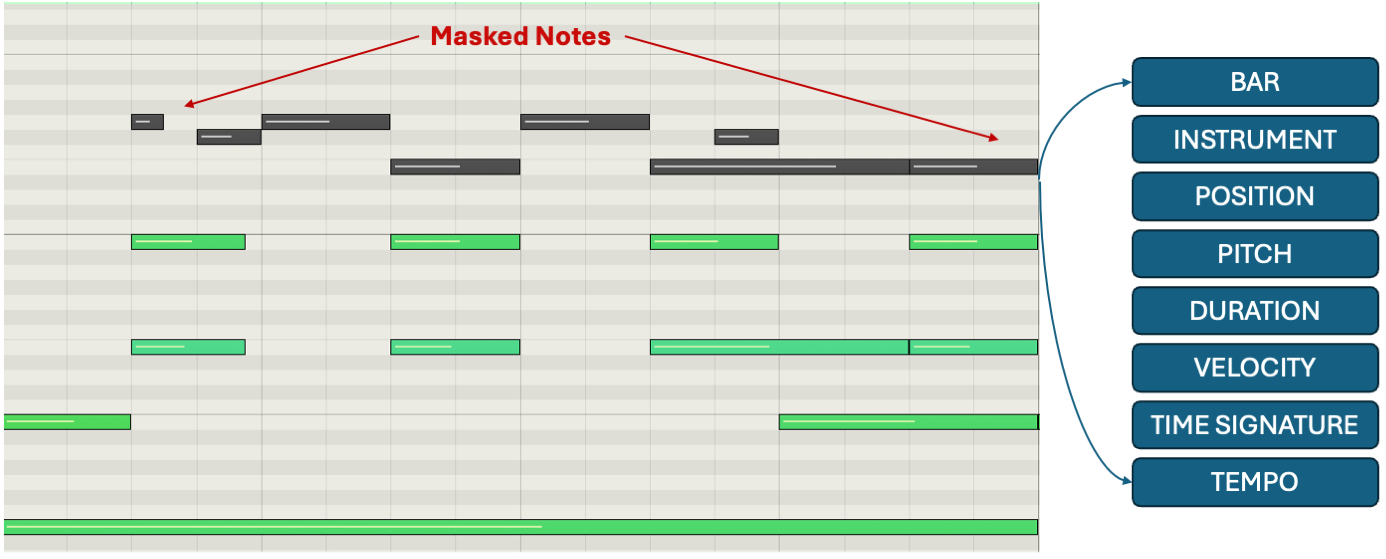}
    \caption{Masking functionality performed by the system. In this diagram, the green notes are unmasked and will remain unchanged by the system. The dark notes are masked and will have one or more of their 8 attributes altered by the AI. The 8 attributes (Bar, Instrument, Position, Pitch, Duration, Velocity, Time Signature and Tempo) are listed to the right.}
    \Description{Provides a representation of how the variation system works. Depicts a piano roll with various notes, some notes are green and others are grey. Grey notes are labelled as being Masked Notes. Next to the piano roll are 8 boxes containing names of attributes the system can alter.}
    \label{fig:MusicBert Example}
\end{figure}

Following these two workshops, a musical variation tool was developed with its main objective to assist in ideation of musical phrases. The current musical variation system utilises MusicBert \cite{zeng2021musicbert} and is based on the MidiFormers project\footnote{https://github.com/tripathiarpan20/midiformers}. MidiFormers utilises masked prediction on a MIDI file to generate remixes of different instruments in a song, employing the MusicBERT model to predict the masked notes. MusicBERT’s Octuple encoding allows for remixing across eight different attributes: Bar, Instrument, Position, Pitch, Velocity, Tempo, and Time Signature. In this encoding, each note is associated with these attributes, and by masking and predicting them, variations in the overall track can be created. An illustration of the masking functionality can be seen in figure \ref{fig:MusicBert Example} and a description of the attributes in Table \ref{tab:attributes}. This concept was selected as a basis for our system for the following reasons: (1) Allows users to control the level of variation by specifying the number of notes to be masked—the more notes masked, the greater the variation. (2) The Octuple encoding method provides inherent control over 8 musical attributes, allowing the user to control not only the level of variation but also the type of variation. These considerations also led us to move away from using Music Transformer, as its lack of inherent controllability \cite{young2022compositional} made it challenging to implement the features desired by participants. Our project built upon the MidiFormers concept and introduced three significant functionality changes to make it better suited for our task. These changes were made to give users more direct control over the model, better addressing the co-design workshop requirements that highlighted the importance of control in the creative process.

\subsubsection{Add New Notes}
While MidiFormers can generate interesting variations of different instruments in a MIDI file, we identified limitations relating to the complexity of the generated variations, particularly on smaller musical sequences. To combat this, we built a new function that provides users the ability to add new notes into the sequence. New notes are introduced as masked tokens, initially assigned only an instrument and bar value. The model then predicts their remaining attributes, such as pitch and position.

\subsubsection{Bar Control}
To give users better control over which parts of a track are varied, we developed a bar specifier function that lets users select specific bars for variation. This allowed users to include extra musical context that the model could use during prediction, without those parts of the track also being varied.

\subsubsection{Bar-Level Masking}
MusicBERT was trained using bar-level masking. This meant selecting a single attribute, masking it over all notes in a bar and then predicting. However, MidiFormers still performs random note masking, leading to sequences that can sound disjointed. In contrast, by applying bar-level masking, MusicBERT can predict connected, continuous note sequences, leading to more cohesive variations. The final version provided an option for both random masking and bar-level masking.

\begin{table*}[]
    \centering
    \begin{tabular}{|c|p{0.7\textwidth}|}
    \hline
    \textbf{Attribute} & \textbf{Description} \\
    \hline 
    Bar     &  Describes which musical bar in the track the note is assigned to. Supporting up to 255 bars of music. \\
    \hline
    Instrument & Describes which instrument in the track plays this note. Supporting 129 different MIDI instruments. \\
    \hline
    Position & Describes the onset of the note in the bar with a granularity of 1/64 note.\\
    \hline
    Pitch & Describes the pitch of the note supporting 128 MIDI pitches. \\
    \hline
    Duration & Describes the length of the note. \\
    \hline
    Velocity & Captures the dynamics and articulations of a note with low velocity notes often sounding quieter and high velocity notes louder. \\
    \hline
    Time Signature & Time signatures denote how many of a particular note should be in the measure of each song. For MusicBERT all notes are assigned a time signature value. \\
    \hline
    Tempo & The tempo of the song at the point of that note. \\
    \hline
    \end{tabular}
    \caption{The various attributes that notes posses in the octuple encoding and a short description of their significance.}
    \Description{Contains control attributes for the music system and a description of these attributes.}
    \label{tab:attributes}
\end{table*}

\subsection{Insights from an Ecological Evaluation}
To obtain intermediate feedback and further design suggestions from musicians, we conducted a two-week ecological evaluation study that provided participants with continuous access to the system within their own creative environments. Ecological evaluation \cite{Brunswik+1956} is a methodology that assesses systems within a participant's own space and time, rather than in a controlled lab environment. This type of methodology is similar to in-the-wild studies \cite{Chamberlain2012,Benford2013} and is widely used in music research \cite{habibi2022music,deruty2022development}. Ecological evaluation is ideal for musicians who have personalised setups at home or in a studio that they are accustomed to using. A total of (n=6) participants took part in this study and used the system for a combined total of 35 hours. Three of the participants had taken part in one of the previous workshops and three participants were new to this study. Participants were provided with a \$30 USD equivalent gift card for their time. Participant demographic information can be seen in Table \ref{tab:w3participants}.

\begin{table}
    \centering
    \resizebox{\columnwidth}{!}{
        \begin{tabular}{|c|c|c|c|}
        \hline
         Participant & Gender & Age & Background  \\
         \hline
         P5  & M & 34-45 & Classical french horn player and musical educator \\
         P8 & F & 25-34 & Background in Classical Piano\\
         P11 & F & 25-34 & Songwriter, Composer and Professional Producer\\
         P12 & M & 18-24 & Musician and songwriter in local band \\
         P13 & M & 25-34 & Guitarist and Songwriter \\
         P14 & M & 18-24 & Electronic Music Composer and Producer \\
         \hline
        \end{tabular}
    }
    \caption{Demographic information of participants who took part in the ecological evaluation.}
    \label{tab:w3participants}
\end{table}

Before gaining access to the system, participants watched a video explaining how to use the system and its various features. The system was hosted on an Amazon Webservice (AWS) EC2 server for a two week period and was accessed using a custom domain. Each participant was provided a unique username and login that ensured their work could be saved on the server. To minimise server load, participants were asked to limit their inputs to approximately 15 bars of music. Beyond this guideline, they were given full freedom in how they could use the system. Participants were provided with prompts to guide their compositions, but they had the option to disregard these and instead focus on their personal or professional projects. After a week of using the system, participants received suggestions on how to use the tool from the project's lead developer. The developers unique understanding of the tool, provided a set of considerations that participants could use if they were having difficulty generating desired outputs. These considerations included which musical attributes provided the best musical variation as well as a method for generating variations that involved chaining variations. However, participants did not have to use these suggestions and could use the tool in any way they preferred. While using the system, participants were asked to keep a journal that they would use to keep track of noteworthy interactions or design suggestions. This ensured that participants would remember useful design suggestions throughout the two-week period. Following this time, participants took part in a focus group aimed at providing an initial assessment for the tool as well as providing design insights into what an initial release of this tool should look like. The journal data and focus group transcripts were analysed using the same method described in section 3.1.

\subsubsection{Understanding the Application of the System}
Participants spent a lot of time \textit{Experimenting} with the system and provided feedback on the \textit{Best Workflows} to utilise with the tool. The \textit{Chaining} technique was favoured by some participants, with P13 noting that "Chaining provided [them] with new chords and bass notes that really inspired some great ideas". Many participants noted the importance of \textit{Filtering and Refining} ideas with P11 stating that "some of the ideas I really liked, but to make them usable, it would just be getting rid of some of the random stuff that it had in there". This process of "choosing what [they] liked and cutting it up" (P14) highlighted that the value of this tool was in "the moments instead of the whole" (P11). As expected, participants observed that the system's impact was most significant in the \textit{Ideation} phase - "it sort of gave me something new I never really thought of before" (P7), "I would never have thought of that kind of progression. So yeah, little bits and pieces out of it could be quite remarkable" (P5). P8 likened their use of the system to writing with ChatGPT \cite{chatgpt2022} noting that "I just choose, particular notes or the melody which I think is really good...it can inspire me in many cases but also leaves me control". When reflecting on their finished compositions, participants often noted that although their music contained minimal output from the system, with the majority of the creative process remaining in the hands of the individual composer, the system inspired them towards the final composition - "So in the end, I guess not a lot of what I made came from rhapsody refiner, but it gave me a cool idea that I turned into something" (P13). In their journal, P11 stated that "I sat down with the intention of working with the tool but the first hour was spent refining the previous ideas it gave me. There was enough there in the midi files that I had direction to work with." This further reinforces the idea that while the system was helpful, the music ultimately relied on the human composer. This sentiment was echoed by other participants, with P5 noting in their journal a comparison between an automatic music generation tool, UDIO\footnote{https://www.udio.com/}, and our system. They remarked, "UDIO has limited creative input – ideas that are then used to generate a full piece" whereas generating smaller fragments with our tool placed "more power in the hands of the composer".

\subsubsection{Design Suggestions from Musicians}

Participants noted that the existing control functionalities offered users an adequate level of control over the generation process - "One thing I like is that the system offers the user a lot of controllability, and it only took a few tries for me to find the combination of settings that best suited my ideas for the current music pieces" (P8). Some controls, such as the add new notes function were particularly liked and were referenced as improving the generated variations "then when I hit, add notes...then it really started generating some interesting stuff" (P13). When discussing the eight musical attributes\footnote{See section 3.3} that users could control, there was some disagreement on the value of certain variables. For example, when referring to the pitch attribute, P12 noted "I've found that the best results were when changing the rhythm rather than the pitch. I think that's where I had the most fun with it.' while other participants identified pitch as one of the primary attributes they altered. P12 suggested that this preference might stem from their background in playing the drums, stating, "Rhythm is something I feel more comfortable working with" emphasising how a musician's \textit{Personal Background} can influence their needs from a co-creative AI. The effect of a musicians \textit{Personal Background} became more apparent when discussing the type of attributes included, with P8 stating that "I found the term position confusing because its not a musical term" and suggesting that "note division within a beat" would be a more intuitive attribute. This confusion was not universal; for example, P14 accurately interpreted the attributes, noting that their familiarity with MIDI and background as an electronic musician likely contributed to their understanding of the attributes. Through this discussion, the importance of \textit{Machine Language vs Music Tradition} became apparent with P5 noting that "it kind of sounds like there's a missing link between the translation between human tradition and how music is being conceived in human minds and human culture". The Time Signature attribute served as a good example of this, with participants noting that in its current form - as an attribute applied to each individual note - lacked significance and that when altering it "nothing would change" (P14). Instead, participants suggested that variations in musical accents would provide a more engaging attribute. Participants also echoed points in workshop 2 regarding DAW integration, highlighting the importance of building co-creative systems within their existing tool sets.

\subsubsection{Outcomes of Model Development and Ecological Study}
Through this two week study we gathered evidence that this tool was useful and that there was a \textit{Desire to Adopt the Tool} amongst participants - "I do see myself using a tool like this in my workflow" (P14) , "I think, having this tool just to give you a new perspective, it's pretty helpful" (P12). Discussions with participants also suggested that this tool and their usage of it through \textit{Filtering and Refining} provided them with \textit{Ownership of the Creative Process}, an important theme from workshop 1. Additionally, the focus groups revealed how an individual's personal musical background influenced their requirements for the system as well as highlighting the disparity between traditional music terminology and machine language.

\section{Design Insights for Building Co-creative AI for Practising Musicians}
After completing all studies, we analysed the various themes and developed a set of design insights that future designers of co-creative musical systems for practising musicians could consider. We believe these insights extend beyond our system and are relevant to various types of musical co-creative AI, not limited to composition tools.

\subsection{AI Collaborator vs AI Tool}
The concept of Human-AI collaboration is gaining traction in creative AI research, with emerging frameworks \cite{Rezwana2023,Muller,Moruzzi2024} and studies \cite{McCormack2020} envisioning a future where artists might create alongside artificial agents. However, in our study, there was resistance from practising musicians when AI was presented as a collaborator in workshop 1. Some expressed strong views, such as P3, who stated, "AI does not have the right to have an opinion," while others emphasised that a shared human experience is crucial for meaningful collaboration in music. These sentiments were tied to the importance of their human embodiment in their practice, with many participants stating that much of their work is focused on capturing and conveying a 'feeling.' Participants in the workshop viewed the term 'collaborator' as almost sacred, believing it should not be used to describe a machine and while these results can not be generalised to all practising musicians, they raise important questions for future designers of co-creative AI such as how is the term 'collaborator' being defined in the context of musical co-creative AI? Should Human-AI collaboration attempt to mimic Human-Human collaboration? Our study suggests that Human-AI collaboration in music should not attempt to replicate Human-Human collaboration, as AI will never fully embody important human qualities - "you actually need empathy, to collaborate, you need that human emotion that the AI can't have". Instead, participants were far more receptive to using AI as a tool and as seen in the ecological study were willing to incorporate our co-creative AI-based tool into their musical practice. Therefore, when designing and developing technology for practising musicians, designers should consider how the framing of their system, as a collaborator or as a tool, may impact its adoption.

\subsection{Ownership of the Creative Process}
Throughout this study, it was clear that participants valued their \textit{Ownership of the Creative Process} and were less interested in tools that automated their music making. They wanted to have full control over the final decisions and confidently claim the finished product as their own. This is unsurprising considering how participants referred to their practice as a "lifetime of perfecting a craft" and often quoted the personal significance of their identity as composers. This finding is consistent with research in other creative fields \cite{oh2018lead}, where participants were reluctant to let AI take the lead. In our study, musicians expressed a strong desire to retain control over elements like "melody and chords" (P5). The variation tool we built was designed to give control of the creative process to users through ideation and was welcomed by participants who stated that there was "more power in the hands of the composer", demonstrating the value of a tool that leaves ownership of the creative process in the hands of the musician. Other research \cite{tchemeube2023evaluating} has also hinted at the value musicians place on owning their creations, further supporting this design insight. Therefore, future designers of co-creative AI should consider the significance of a practising musician's identity and their creative drive, and develop technology that supports their practice rather than replacing key aspects they value.

\subsection{Traditional Music Terminology vs Machine Representation}
In our study, some participants raised concerns about the naming and types of attributes they had control over. It became clear that there were translation issues between traditional musical terminology and the machine language derived from the MIDI format. While this was not a concern for all participants, some existing methods for encoding musical data in models \cite{huang2018music,zeng2021musicbert} do diverge from the terminology that many musicians are familiar with - potentially making it more challenging for musicians with a traditional background to control them. Designers should therefore be mindful of this and work to bridge the gap between traditional music terminology and machine representations, particularly when designing for musicians with a traditional background. For example. in our study, participants suggested to use a note 'division' attribute, rather than a note 'duration' and 'position' attribute, noting this would be more intuitive to their practice. 

\subsection{Influence of Diverse Musical Practices on Design}
A key strength of this study was the range of musical backgrounds represented among the participants. Their diverse experiences, from formally trained composers to local artists and songwriters, improved our understanding of their needs concerning co-creative AI. Importantly, their varied backgrounds highlighted that different groups of musicians have distinct requirements. For example, P14, an electronic music composer, found the control attributes 'position' and 'duration' intuitive, whereas P11, who had a more traditional background, found these terms confusing and less meaningful. Designers should therefore be mindful of the specific group of musicians they are designing for and actively involve them in the design process to gain a better understanding of their needs and requirements, recognising that what works for one group may not be suitable for another.

\subsection{Integration into Current Ecosystem}
Throughout the study, participants noted the importance of incorporating these technologies into their DAWs. For those experienced in music production, this insight may seem obvious and efforts are already underway to incorporate music models into widely used DAWs \cite{tchemeube2023evaluating,banar2023tool,roberts2019magenta}. Although integrating our system into a DAW was beyond the scope of this study, participants made it clear that future designers should consider from the outset how their system will fit into the existing musical ecosystem. For example, when working with large deep learning models, designers should consider the increased time and resource demands needed to run these systems on personal laptops versus the GPU-powered machines they were developed on. Without these considerations, their technology is unlikely to be adopted by practising musicians.

\section{Discussion}
\subsection{Reflecting on the Impact of Our Research}

\begin{quote}
    "On a whole I really enjoyed working with the system and I found that it made me compose in a way that I wouldn’t normally. There were sounds in the compositions, more specifically colours of chords that I wouldn’t identify with my own style, I think this was the best part of it and I can see real benefit for this alone. I do notice that composing can become routine and habitual in that I often don’t stray away from what I know will work. So the nice thing about this was it forced me out of those habits and it was a more curious and open ended style of composition. As I write this it also makes me realise that all of my previous comments coming from the mentality of trying to get the tool to work with what is my already established composition practice and though I can imagine ways that would be possible and beneficial, I see upon greater reflection that the most salient benefit was forcing me to relinquish some of that control to make space for something new. That’s pretty special I think."
    \flushright --- Excerpt From P11's Journal
\end{quote}

As researchers, we must remain mindful that the technology we create can disrupt industries and impact lives, with no certainty that it will add value or lead society toward a better future \cite{koepsell2010genies,Coiera2011}. Therefore, it's important that we try to understand our users and their industry when designing new technology, ensuring that our work is both meaningful and justified. In this study, participants expressed strong opposition to the idea of collaborating  with an AI. Although some may view these concerns as extreme, they reflect genuine fears of losing identity in the music world through the rise of automated composers. However, these participants were not completely against using co-creative AI in music, describing a desire to use these systems as tools - tools that ultimately leave ownership of the creative process to them. We drew on their guidance to develop a prototype system that participants were eager to integrate into their practice, — a result we attribute to the workshops, which provided participants with the opportunity to express their preferences and influence the system's design. However, while Creative AI has potential to add to musical practice, we must still remain mindful during its development to avoid the risk of damaging a deeply human practice, one which has held significance for various cultures throughout our history.

\subsection{Co-Design in Developing Co-creative Musical AI}
Co-design is a well-established methodology that has been successfully applied in various fields \cite{Weitz2024,Nicholson,Harrington2018} and enables researchers to gain direct insights into the needs of their user groups throughout development. As noted in Section 2, co-design is rarely applied in the development of co-creative musical AI and while there are plenty of impressive musical AI tools \cite{MusicVAE,tchemeube2023evaluating,thickstunanticipatory,huang2018music}, designers and researchers could still benefit from incorporating musicians into the design process. Through our co-design study, we identified insights that aligned with the goals of MidiFormers and Music Transformer, such as the importance of controllability and generation quality. However, our co-design approach also revealed broader considerations, including the importance of preserving musicians' ownership of the creative process. This led us to extend MidiFormers beyond its original scope, transforming it into an ideation tool that was received well by participants. Furthermore, in this study, it was clear that different types of musicians have different needs and that a tool built for a classical composer could likely have different requirements than a tool built for an electronic musician. Through a co-design methodology, researchers could more effectively ensure that they are addressing the specific needs of different groups of musicians, as done in Louie et al. (\citeyear{ryan2020})\cite{ryan2020} who worked with novice musicians.

It is worth acknowledging that while we consider co-designing with practising musicians ideal, it can be difficult to find professional/practising musicians that are willing to take part in these studies, particularly longer evaluations such as our two-week ecological study. However, we hope that this work will encourage other researchers to use a co-design methodology when developing co-creative AI and recognise the benefit of including practising musicians throughout their design process.

\subsection{Study Limitations \& Future Work}
One limitation regarding our workshops was the lack of representation from older musicians, with the oldest age bracket being 34-45. This means we may have missed some important design considerations for that specific group of users. However, we argue that this provides further motivation for using a co-design methodology when designing co-creative musical AI. Through this methodology, researchers can understand what this subgroup of musicians would need from these tools by incorporating them into their design process. Providing not just a system usable by the target groups, but also valuable insights into the general needs of that target group.

Due to unexpected funding constraints, the remuneration we could provide participants in the ecological study and second workshop, was substantially less compared to the first workshop, which ultimately hindered our recruitment strategy. Despite this, we still found there to be interest in the study, with many participants interested in the opportunity to have an impact on the design of AI music tools, particularly with AI becoming more prominent in their field.

While out-of-scope for this project, we hope to build our variation system into a plugin that can be added to popular DAWs such as Abelton and Logic. This would enable study participants to continue using the tool in their practice while also making it accessible to a wider variety of practising musicians. Additionally, we aim to improve our system and use it to understand how variation generation can be used as a mode of interaction within co-creative AI music composition.

\section{Conclusion}
As AI becomes more prevalent in the music industry, it is important that we are designing tools that meet musicians real needs, ensuring the technology we create has a positive impact on the industry. In this paper, we presented a case study that used a co-design methodology with 13 practising musicians to develop a co-creative AI, demonstrating how this approach could define a tool’s role in musical practice early in development. Through this approach, we also identified various design insights that may be useful for future developers, such as the impact of framing a system as a collaborator or tool, the importance of owning the creative process, and how diverse musical backgrounds shape user needs.

\begin{acks}
This research was supported by an Australian Government Research Training Program (RTP) Scholarship.
\end{acks}

\bibliographystyle{ACM-Reference-Format}
\bibliography{sample-base}


\end{document}